\documentclass[twoside]{article}
\usepackage{fleqn,espcrc2,epsf}

\newcommand{\AmS}{{\protect\the\textfont2
  A\kern-.1667em\lower.5ex\hbox{M}\kern-.125emS}}

\def\LL{\left\langle}	
\def\RR{\right\rangle}	
\def\LP{\left(}		
\def\RP{\right)}	

\def\BE{\begin{equation}}
\def\EE{\end{equation}}
\def\BEA{\begin{eqnarray}}
\def\EEA{\end{eqnarray}}
\def\EL{\nonumber\\}

\title{Determining hybrid content of heavy quarkonia using lattice
nonrelativistic QCD}

\author{Tommy Burch,
	Kostas Orginos\thanks{Present address: RIKEN-BNL Research Center,
			      Bldg 510a, Upton, NY 11973-5000},
	and Doug Toussaint\address{Department of Physics,
				   University of Arizona, \\
		 		   Tucson, AZ 85721}}
       
\begin{document}

\begin{abstract}
Using lowest-order lattice NRQCD to create heavy meson propagators and
applying the spin-dependent interaction,
$c_B^{} \frac{-g}{2m_q}\vec\sigma\cdot\vec{B}$, at varying intermediate
time slices, we compute the off-diagonal matrix element of the Hamiltonian
for the quarkonium-hybrid two-state system. Diagonalizing this two-state
Hamiltonian, the admixture of hybrid ($|Q\bar{Q}g\rangle$) in the ground
state is found. We present results from a set of quenched lattices with an
interpolation in quark mass to match the bottomonium spectrum.
\end{abstract}

\maketitle

\section{INTRODUCTION}

While experimental confirmation of the existence of hybrid states --
hadrons with a gluonic excitation ($q\bar qg$, $qqqg$, etc.) --
remains elusive, much has been done theoretically, both on and off the
lattice, to determine possible consequences of their existence.
In particular, mixing of (non-exotic) hybrids with normal hadronic states
(with the same $J^{PC}$) is expected to occur, necessitating a redefinition
of the true hadronic ground state in terms of its constituents: e.g.,
\BE
|\Upsilon\rangle = A_S |b\bar b\rangle + A_H |b\bar bg\rangle + ....
\EE
We offer the bottomonium example for a reason: it is this system which we
study with our simulations.


Since this system is composed of relatively heavy quarks,
we use the nonrelativistic approximation to QCD (NRQCD)
\cite{Le92} when we perform the lattice simulations.
NRQCD has been used before to study heavy hybrids \cite{NRQCD_HYBRIDS}. The
main difference in the current study is the fact that we apply the
lowest-order (in $\frac{1}{m_q}$) spin-dependent interaction
($c_B^{}\frac{-g}{2m_q}\vec\sigma\cdot\vec B$) ``perturbatively,'' i.e., at
a single time slice. Then, using different source and sink operators
(quarkonium $\rightarrow$ hybrid, and {\it vice versa}), we are able to
extract the off-diagonal matrix element of the Hamiltonian for this
quarkonium-hybrid two-state system. We then diagonalize this new
Hamiltonian and find the admixture of hybrid within the true ground state
\cite{BuOrTo01}.

\section{METHOD}

In our implementation of NRQCD,
we use a time-step-symmetric form of the heavy quark evolution operator
\cite{Le92}:
\BEA
\lefteqn{G(\vec{x}, t+a) =}\EL
&&\LP 1 - \frac{aH_0}{2n}\RP^n U^\dagger_t(x)
\LP 1 - \frac{aH_0}{2n}\RP^n\EL
&&\times\ (1 - \delta_{t',t}^{} a\delta H) G(\vec{x}, t),
\EEA
with a value of $n=2$ (more than sufficient for stability,
$n>\frac{3}{2m_qa}$).
For simplicity, we use only the lowest-order term of the heavy quark
expansion in the diagonal part of our Hamiltonian:
\BE
H_0 = \frac{-\vec{D}^2}{2m_q},
\EE
where $\vec D$ is the covariant derivative. The spin-dependent term appears
in the interaction
\BE
\delta H = c_B^{}\frac{-g}{2m_q}\vec\sigma\cdot\vec{B},
\EE
which is applied only at a single, intermediate time slice, $t'$.

\begin{table}[hbt]
\caption{
Meson operators.
}
\label{OPERATORS}
\begin{tabular}{lc}
\hline
$J^{PC}$ & Operator \\ \hline
$0^{-+}$ S-wave & $\bar{Q} Q$ \\
$1^{--}$ S-wave & $\bar{Q}\sigma_i Q$ \\
$0^{++}$ P-wave & $\bar{Q}\sigma_i D_i Q$ \\
$1^{++}$ P-wave & $\bar{Q}\varepsilon_{ijk}\sigma_j D_k Q$ \\
$2^{++}$ P-wave & $\bar{Q}(\sigma_i D_j + \sigma_j D_i - \frac{2}{3}\delta_{ij}\sigma_k D_k) Q$ \\
$0^{-+}$ Hybrid & $\bar{Q} \sigma_i B_i Q$ \\
$1^{--}$ Hybrid & $\bar{Q} B_i Q$ \\
\hline
\end{tabular}
\end{table}

For our lattice mesons, we use an incoherent sum of point sources:
at the source end, we start
with a given quark color and spin at all spatial points, without fixing the
gauge; at the sink end, we sum over all the contributions where the quark
and anti-quark are at the same spatial point. Since the lattices are not
gauge-fixed, we expect the contributions from sources with the quark and
anti-quark at different spatial points to average to zero.
We combine the quark and anti-quark sources (propagators) with the
appropriate spin matrices and gauge fields to construct the meson operators
at the source (and sink) time slices. The meson operators we use are
displayed in Table~\ref{OPERATORS}.

Using identical source and sink operators, ``unmixed'' meson (and hybrid)
propagators are constructed. We use the following form to fit the resulting
correlators:
\BE
C(t) = A_0 e^{-m_0^{}t} + A_1 e^{-m_1^{}t}.
\EE
For the ``mixed'' propagators, the source and sink operators differ
(quarkonium $\rightarrow$ hybrid, and {\it vice versa}). These are also fit
with a two-exponential form. However, since the source and sink operators
are different, and since there is the single interaction ($\delta H$) at the
intermediate time slice ($t'$) causing the mixing between these
configurations, the first exponential in this fit can be expressed as
\BEA
\lefteqn{C_{mix}(t) =}\EL
&& A_{0,src}^{1/2}A_{0,snk}^{1/2}
\LL 1\mbox{H}\left|
c_B^{}\frac{-g}{2m_q}\vec\sigma\cdot\vec{B}\right|1\mbox{S}\RR \EL
&&\times\ e^{-m_{0,src}^{}t'} e^{-m_{0,snk}^{}(t-t')} + ....
\EEA
Knowing the amplitudes and masses from the unmixed propagators and fitting
this correlator in the region $t>t'$, we extract
the off-diagonal matrix element. We then repeat this calculation for larger
and larger values of $t'$ in search of a plateau, indicating the decay of
excited-state contributions (e.g., from $|2$S$\rangle$ or $|2$H$\rangle$,
depending on the source) to the mixed amplitude.

\section{RESULTS}

The heavy meson propagators are evaluated on a set of quenched, improved
lattices (100 configurations with Symanzik 1-loop-improved gauge action,
$20^3\times 64$, $\beta = 8.0$; see Ref.~\cite{MILC_SCALING}). The lattice
spacing is determined using the spin-averaged 1S-1P mass splitting for
bottomonium; $a^{-1} =$ (440 MeV)$/(a \Delta M_{SP}^{}) = 1590(30)$
MeV. We also construct non-zero-momentum $1^{--}$ S-wave
propagators and use the resulting dispersion relation to determine the
kinetic mass for this meson. This kinetic mass is calculated for two input
values of the quark mass ($m_qa = 2.5$ and 2.8) and an interpolation is
then performed to match this meson mass to the experimental
value for the mass of the $\Upsilon$ (9.46 GeV). We thus arrive at a
(lattice-regularized) physical value for the mass of the bottom quark
($m_b^{}a = 2.71$).

\begin{figure}[htb]
\vspace{-5mm}
\epsfxsize=3in
\epsfysize=3in
\epsfbox[0 0 4096 4096]{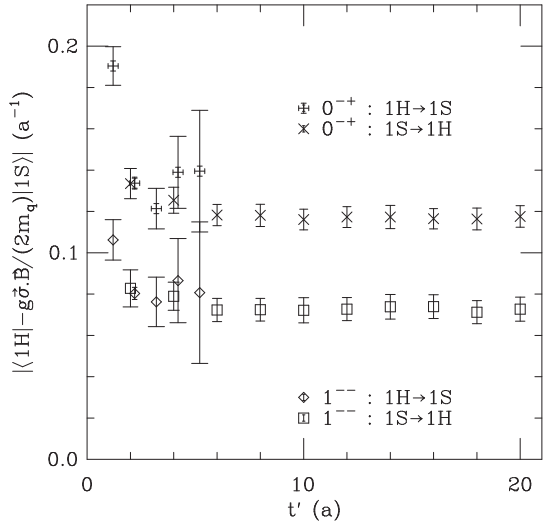}
\vspace{-1cm}
\caption{
Magnitude of the off-diagonal matrix element of the Hamiltonian (in lattice
units) for the S-wave / Hybrid two-state system vs. the time slice, $t'$,
at which the interaction ($\delta H$) is applied ($m_qa=2.5$).
}
\label{SIGMABFIG}
\end{figure}

\begin{figure}[htb]
\vspace{-5mm}
\epsfxsize=3in
\epsfysize=3in
\epsfbox[0 0 4096 4096]{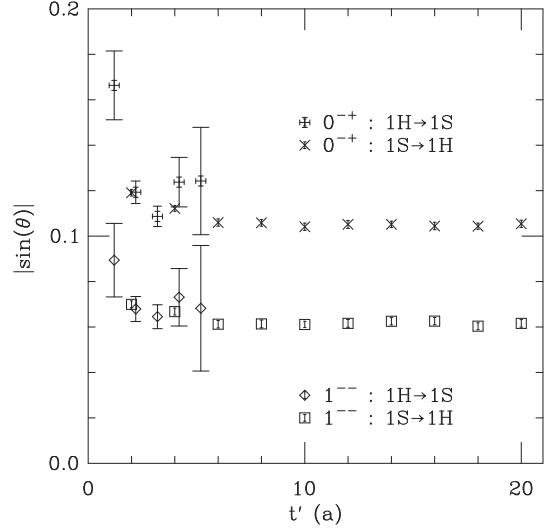}
\vspace{-1cm}
\caption{
Mixing angle, $\sin(\theta)$, vs. the time slice, $t'$, at which the
interaction term ($\delta H$) is applied ($m_qa=2.5$).
}
\label{SINSFIG}
\end{figure}

Appearing in Fig.~\ref{SIGMABFIG} are the results for the off-diagonal
matrix element, plotted as a function of the interaction time slice, $t'$.
After diagonalization of the Hamiltonian, the hybrid content of the ground
state is found and expressed in terms of a mixing angle ($\sin\theta$). A
plot of this quantity versus $t'$ is shown in Fig.~\ref{SINSFIG}. As can be
seen in both of these figures, a plateau is reached at rather low values
of $t'$. Using the values at $t'=10$ ($\chi^2/d.o.f. < 1$) and interpolating
in the quark mass to the physical value, we find the following
configurations for the $\Upsilon$ and $\eta_b$ ground states:
\BEA
\lefteqn{|\Upsilon\rangle = 0.99837(6)\ |1\mbox{S}(1^{--})\rangle}\EL
&&-\ 0.057(1)\ |1\mbox{H}(1^{--})\rangle ,
\EEA
\BEA
\lefteqn{|\eta_b^{}\rangle = 0.9953(1)\ |1\mbox{S}(0^{-+})\rangle}\EL
&&+\ 0.097(2)\ |1\mbox{H}(0^{-+})\rangle ,
\EEA
with the tree-level value $c_B^{}=1$.

In an attempt to determine $c_B^{}$ nonperturbatively, we perform
additional runs with the interaction, $\delta H$, applied at {\it all}
intermediate time slices and with the values of $c_B^{}=1$ and 2. We then
use the $\Upsilon-\eta_b^{}$ mass difference (a quantity having a quadratic
dependence upon the $\vec\sigma\cdot\vec B$ term) to set the value of
$c_B^2$. Unfortunately, as there is no experimental value for the mass of
the $\eta_b^{}$, we rely on potential model results \cite{POT_MODELS} which
put this mass difference in the $30-60$ MeV range. Combining this with our
lattice results for the mass difference --
$\Delta M_{\Upsilon-\eta_b}^{}=18.4(4)$ and 71.3(2.1) MeV for $c_B^{}=1$
and 2, respectively -- suggests $c_B^2 \sim 1.7-3.4$. The resulting
probability admixtures of hybrids within the bottomonium ground states thus
become
\BE
|\langle 1\mbox{H}|\Upsilon\rangle|^2 \sim 0.0054 - 0.011 ,
\EE
\BE
|\langle 1\mbox{H}|\eta_b^{}\rangle|^2 \sim 0.016 - 0.032 .
\EE
Obviously, the uncertainty is dominated by the lack of a reliable value
for the $\Upsilon-\eta_b^{}$ mass difference.
Note the factor of $\sim 3$ enhancement for the mixing in the $0^{-+}$
channel, due to spin statistics \cite{Ba01}. Interested readers can find
a more detailed discussion of the results for the $1^{--}$ channel in
Ref.~\cite{BuOrTo01}.

The simulations are to be repeated on lattices with a different value of
the coupling in order to attempt a continuum extrapolation.

\end{document}